# Astrophysical $\gamma$-ray and Neutrino Production Following from the Physics of Photon-Nucleon Interactions

Floyd W. Stecker[1,2]

[1] *Astrophysics Science Division*
*NASA/Goddard Space Flight Center*
*Greenbelt, MD 20771, USA*
[2] *Department of Physics and Astronomy*
*University of California, Los Angeles*
*Los Angeles, CA 90095-1547*

## ABSTRACT

The astrophysical implications of particle photoproduction interactions involving nuclei is considered here, based on the most recent empirical data on particle photoproduction interactions off protons. The implications of photoproduction with helium nuclei are also discussed and compared with $\gamma p$.interactions. It is found that $\gamma$-He interactions, assuming the cosmological abundance of He, produce approximately 10% of the pions as compared with $\gamma p$ interactions. In addition to the production of pions, we also discuss the relative effect of excited nucleon $p\pi$, $p\pi\pi$ resonances and $\rho$,$\eta$, $\omega$ and $K$ production and decay channels leading to neutrino and $\gamma$-ray production. The production of mesons other than pions is found not to be significant for astrophysical calculations of $\gamma$-ray and neutrino production. It is further shown that, for astrophysical purposes, the $\Delta(1232)$ resonance channel clearly dominates. However, a blend of $N^*$ resonances at $\sim$1400 GeV can contribute as much as 20% to the neutrino flux.

## 1. INTRODUCTION

Astrophysical processes that lead to the production of photopions have been recognized for a considerable period and have been discussed in numerous prior studies (Greisen 1966; Zatsepin & Kuz'min 1966; Stecker 1968; Berezinsky & Zatsepin 1970; Stecker 1973, 1979; Kelner & Aharonian 2008; Hümmer et al. 2010; Stecker 2023a,b). In particular, photoproduction processes that lead to $\gamma$-rays and neutrinos in the vicinity of astrophysical objects containing black holes are of significant interest (Stecker et al. 1991; Murase & Stecker 2023), as is the production of ultrahigh energy neutrinos from interactions of ultrahigh energy cosmic rays with photons of the cosmic background radiation (Stecker 1979).

In 2013 the first announcement of high energy astrophysical neutrinos was published by the IceCube Collaboration (IceCube Collaboration 2013). Since then large numbers of astrophysical neutrinos have been detected by the IceCube neutrino telescope (Halzen & Kheirandish 2023). A few $\gamma$-ray sources have also been identified as neutrino sources (IceCube Collaboration 2018, 2022). However, in general, individual $\gamma$-ray sources do not correlate with sources of neutrinos. Thus at present the vast majority of astrophysical neutrinos are of unknown origin. This has lead to the hypothesis that most neutrino sources are "hidden" (Stecker 2013; Murase et al. 2016).

Such recent developments have provided a motivation to carefully reexamine the particle physics processes involved in high energy $\gamma p$ interactions leading to significant $\nu$ and $\gamma$-ray production off nuclei. To this end it is important to use an extensive, up-to-date collection of accelerator data on the production cross sections for $\gamma p$ channels that can lead to astrophycially significant $\nu$ and $\gamma$-ray production.

Previous work on photoproduction since 2000 has relied on the SOPHIA code (Mücke et al. 2000). That code was based on various assumptions to make up for the lack of quality data being used at the time. Among other conclusions, the SOPHIA code predicts a significant amount of $\gamma$-ray production from interactions of higher energy than those producing the $\Delta(1232)$ resonance. Many other works have also relied upon Monte Carlo simulation codes such as GEANT4. Such simulations are optimized for higher accelerator energies, as are other existing numerical packages.

On the contrary, the approach here is to directly utilize and to rely on the presently available data, which are much more reliable and direct than numerical simulations. This approach allows for a more accurate treatment for considering the $\gamma$-ray and neutrino spectra fluxes resulting from astrophysical sources and processes.

The present study makes use of a library of detailed energy dependent cross section data for a large number of production channels that either were not available at the time when the SOPHIA code was written or not used at the time. This library includes all of the particle production channels resulting from $\gamma p$ interactions, including empirical cross section data obtained from both cloud chamber experiments and particle accelerator experiments. The result of this approach will be a reliable determination of the relative significance of the particle production channels involved.

## 2. DIFFERENCES BETWEEN $\gamma$p AND pp INTERACTIONS.

Because of the electromagnetic nature of the photon, interactions with nucleons are of a different nature than that of $pp$ interactions (Chew & Low 1956; Chew et al. 1957). Empirically, as will be discussed, photon-nucleon interactions typically involve the generation of one particle or resonance. (See e.g., Drechsel & Tiator (1992)). Such a particle or resonance can decay in a chain of two-body decays. This process can result in the production of one, two, or three pions. The charged pions then decay leading to neutrino production. The neutral pions decay into $\gamma$-rays.

Nucleon-nucleon interactions are of a strong-interaction, QCD nature. Quantum chromodynamics interactions among the various quarks and gluons (partons) that make up the nucleons determine the production of observed particles in such interactions (Geiger 1993). Nucleon-nucleon interactions can therefore generate a multiplicity of particles, mainly pions, whose number rises with energy, e.g., Stecker (1973). The $pp$ cross section is typically two-to-three orders of magnitude higher than the $\gamma p$ cross section, except in the case of $\Delta(1232)$ resonance production, as we shall discuss.

Various theoretical models have been proposed to describe and explain $\gamma p$ interactions in terms of a hadronic nature for the photon. For example, in one such model the photon can convert into a virtual $\rho$ meson that interacts with partons in the nucleon (Gevorkyan 2019). Since the first vertex of photoproduction interactions is a QED vertex (see, e.g., Fig.3), in all such models the interaction cross section is reduced by a factor of order $\alpha = (e^2/\hbar c) = 1/137$ or more, as compared to the $pp$ cross section.

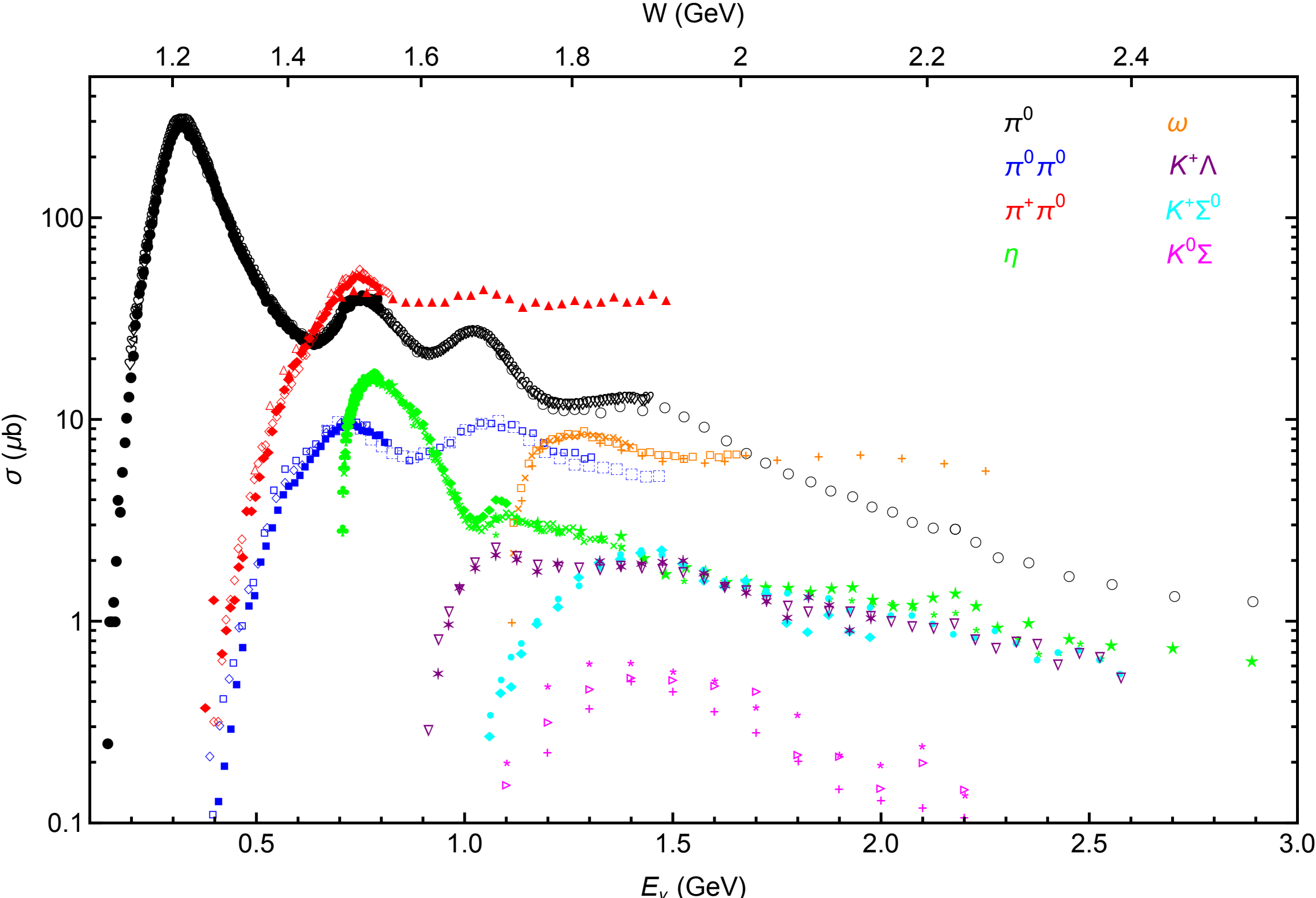

**Figure 1.** Photoabsorption cross sections for the $\gamma p$ channels leading to neutral pions and other $\gamma$-producing particles. The lower scale shows the photon energy in the laboratory system where the target proton is at rest. The upper scale shows the Lorentz invariant cms energy, W, as given by equation (3). Single $\pi^0$ production is shown in black. The data are from Genzel et al. (1973) (right pointing open triangles), Härter et al. (1997) (small solid circles), the CB-ELSA Collaboration et al. (2005) (open circles), Schumann et al. (2010) (solid circles), and the A2 Collaboration at MAMI et al. (2015a) (open hearts). The $\pi^+\pi^0 n$ cross section is indicated in red. The data are from Braghieri et al. (1995) (open upward triangles), Langgärtner et al. (2001) (open diamonds), Ahrens et al. (2003) (solid diamonds), Bartalini et al. (2008) (solid upward triangles), and Zehr et al. (2012) (solid downward triangles). Double $\pi^0$ production is shown in blue with data from Assafiri et al. (2003) (dashed open squares), Thoma et al. (2008) (open squares), Sarantsev et al. (2008) (closed squares), and Schumann et al. (2010) (open diamonds). The $\omega$ cross section data are shown in orange and are from Barth et al. (2003) (+'s), the CLAS Collaboration et al. (2009) (open squares), and the A2 Collaboration at MAMI et al. (2015b) (x's). The $\eta$ cross section data are shown in green and are from Krusche et al. (1995) (open circles), Renard et al. (2002) (closed circles), and Bartalini et al. (2008) (small closed circles). $K^+\Lambda$ production is shown in purple with data from Tran et al. (1998) (large *'s) and Glander et al. (2004) (open inverted triangles). $K^+\Sigma^0$ production is shown in cyan with data from Tran et al. (1998) (closed diamonds) and Glander et al. (2004) (closed small circles). $K^0\Sigma$ production is shown in magenta with data from Klein (2005) (+'s), Lawall et al. (2005) (*'s), and Castelijns et al. (2008) (open right-pointing triangles).

## 3. KINEMATICS

The threshold energy for production of one or more particles in $\gamma p$ interactions by $\gamma$-rays of lab energy $\epsilon$ interacting with protons at rest is given by

$$\epsilon_{th} = \Delta M(\left(1 + \frac{\Delta M}{2M_{\text{target}}}\right) \tag{1}$$

where the quantity $\Delta M$ represents the difference between the mass or masses of the particle(s) produced in the interaction and the mass of the target particle (Stecker 1968, 1971). It follows from equation (1) that the threshold energy for producing a neutral pion is 145 MeV; for a charged pion it is 150 MeV. The energy required for producing an $\eta$ meson is 708 MeV; that for producing a $\rho$ meson is 1086 MeV and for an $\omega$ meson it is 1110 MeV, etc.

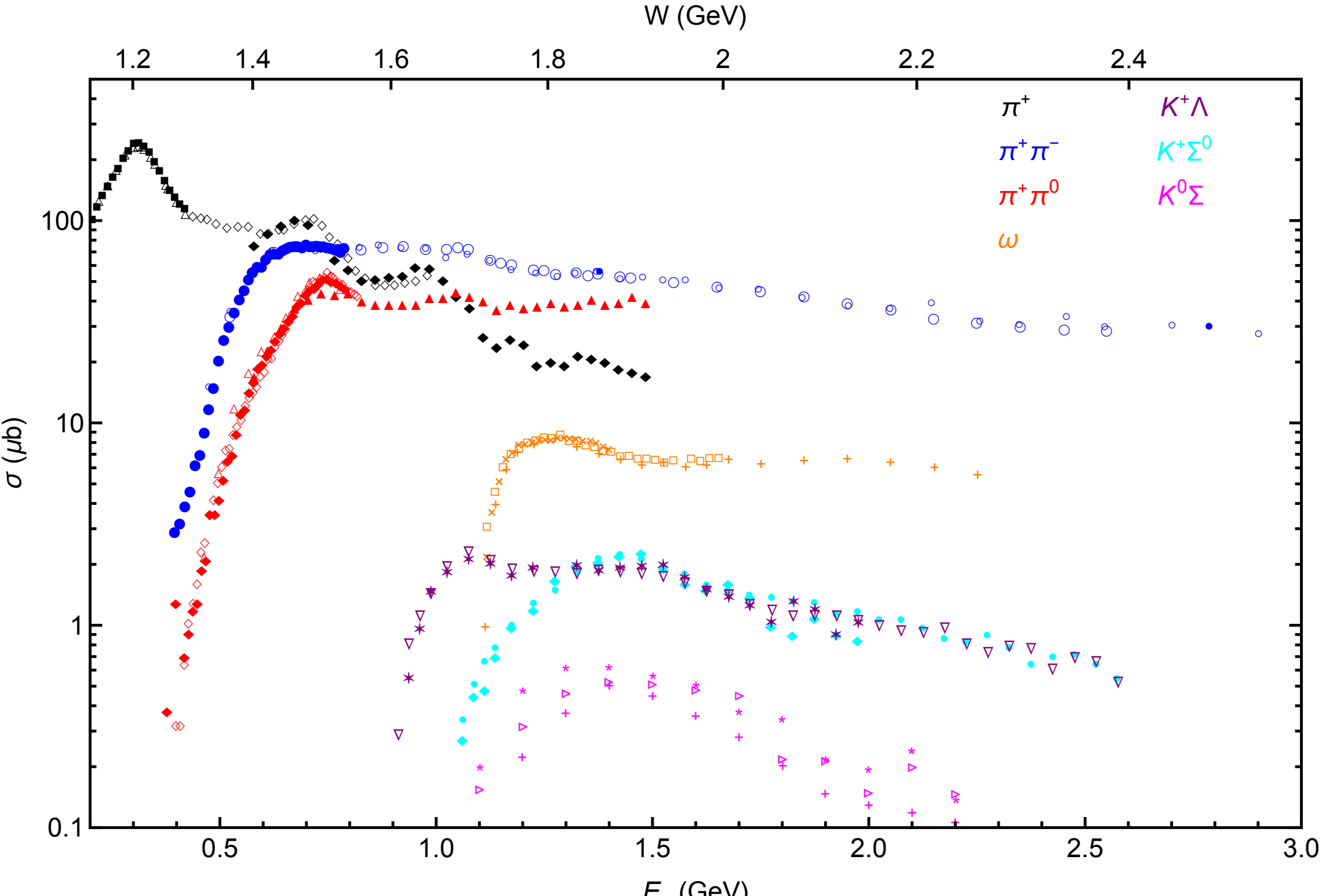

**Figure 2.** Photoabsorption cross sections for the $\gamma p$ channels leading to charged pion and other $\nu$-producing particles. The scales are those as in Fig.1. Single $\pi^+$ production is in black with data from Genzel et al. (1973) (open upward triangles), Büchler et al. (1994) (open diamonds), MacCormick et al. (1996) (solid squares), and Bartalini et al. (2008) (solid diamonds). Double pion $\pi^+\pi^-$ data is shown in blue with data from the Aachen-Berlin-Bonn-Hamburg-Heidelberg-München Collaboration (1968) (small open circles), Ballam et al. (1972) (small solid circles), Braghieri et al. (1995) (solid circles), and Wu et al. (2005) (open circles). Data references for $\pi^+\pi^0$, $\omega$, $K^+\Lambda$, $K^+\Sigma^0$, and $K^0\Sigma$ can be found in the caption of Fig. 1.

The situation is somewhat different for the production of a strange particle such as the kaon. The explanation of the relatively long lifetime of these particles is the conservation of strangeness in strong interactions, first explained by Gell-Mann (1953), who called them "strange", and also by Nakano & Nishijima (1953). The decays of strange particles into non-strange particles violates this conservation law, so that the decays must be weak, thereby accounting for their longevity.

Because of conservation of strangeness, since the $\gamma p$ system has a strangeness number of 0, any strange particles produced in a $\gamma p$ interaction must be produced in associated pairs of equal and opposite strangeness. Such interactions, viz., the production of $\Lambda K^+$, $\Sigma^0 K^+$ and $\Sigma^+ K^0$ pairs have been studied and we have plotted the measured cross sections for these processes in Figure 1. We note that, unsurprisingly, the production of strange particles is insignificant compared with that of non-strange particles.

Almost all decay chains that finally produce pions (and thus $\gamma$-rays and neutrinos are the of result of chains of two body decays. This simplifies the kinematics considerably.

If we label the particles produced $a$ and $b$, then the energies of the particles are exactly determined through conservation of energy and momentum. In the rest frame of the decaying particle of mass $M$ they and are given by

$$E^*_{a,b} = E^*_{p_i} \left( \frac{M^2 + m_{a,b}^2 - {m_{b,a}}^2}{2M} \right). \tag{2}$$

(see Stecker 1971, chap. 1).

For our analysis, we start with the measured cross section data for a $\gamma$-ray of laboratory energy $\epsilon$ interacting with target protons at rest in the laboratory system. The total energy of the system as measured in its center of mass, will be denoted by $W$. The value of $W$ in the photon-proton interaction system is given by

$$W = \sqrt{s} = (M_p^2 + 2M_p\epsilon)^{1/2} \tag{3}$$

The variable $W$ is equal to $\sqrt{s}$, where the Mandlestam variable $s$ (and therefore $W$) is Lorentz invariant, i.e., the square of the total four-momentum of the photon-proton system (Stecker 1968).

We can make use of the Lorentz invariance of $s$ in order to transform to a system where the proton is *not* at rest, but is *relativistic* in the rest frame of an astrophysical source.

In that case, $s$, is given by

$$s = (\epsilon^* + E^*_{p_i})^2 = m_p^2 + 2E_p\epsilon\beta(1 - \cos\theta), \tag{4}$$

where $\epsilon$ and $E_p$ denote the photon and proton energy in the laboratory system, and $\theta$ is the angle between the relativistic proton ($\beta \simeq 1$) and a photon in the lab system. Quantities in the center-of-momentum system (cms) are denoted by an asterisk. For an isotropic photon source we take $< \cos\theta = 0 >$.

Thus, e.g., for a pion produced by two-body decay according to equation (2), its energy in the lab system would be given by

$$E_\pi = (E_p/W)\beta(1 - \cos\theta)E^*_\pi \simeq (E_p/W)E^*_\pi, \tag{5}$$

where we take $\beta \simeq 1$ for the relativistic proton and we take $< \cos\theta >= 0$, assuming isotropic or quasi-isotropic decay in the cms of the interaction.

For a two-body decay of a resonance of mass $M_R \simeq W$, with the energy of the pion and nucleon products given by equation(2), the energy of the secondary pion is given by

$$E^*_\pi = E^*_{p_i} \left( \frac{W^2 + m_\pi^2 - {m_n}^2}{2W} \right). \tag{6}$$

## 4. CROSS SECTIONS

Figure 1 shows the cross section data for channels leading to the production of neutral pions and, ultimately, $\gamma$-rays . The data is shown from experiments where high energy $\gamma$-rays interact with protons at rest in the laboratory system.

Figure 2 shows the corresponding cross section data for channels lead to the production of charged pions and, ultimately, neutrinos.

### 4.1. *Single pion production*

At low laboratory photon energies the single pion production channels (see Section on single pion production) are the most important ones. It follows from equation (1) that the threshold energy for photopion production is $\sim 0.145$ GeV (Stecker 1968). At photon energies 0.145 GeV $\epsilon' \leq 1$ GeV, $\pi^+$ production is dominant. In the energy range around 1.2 GeV GeV, the $\Delta^+_{3/2}$ resonance dominates the cross section.

The existence of the $\Delta^+_{3/2}$ resonance (or isobar) has been known since the 1950s Wilson (1958). It follows from isospin conservation that the $\Delta^+_{3/2}$ decays 2/3 of the time into $p\pi^0$ and 1/3 of the time into $n\pi^+$, with the single pion decay modes having a total branching ratio of 99.4 %. Thus, in this energy range $\pi^0$ production dominates over $\pi^+$ production. This interaction is diagrammed in Fig. 3.

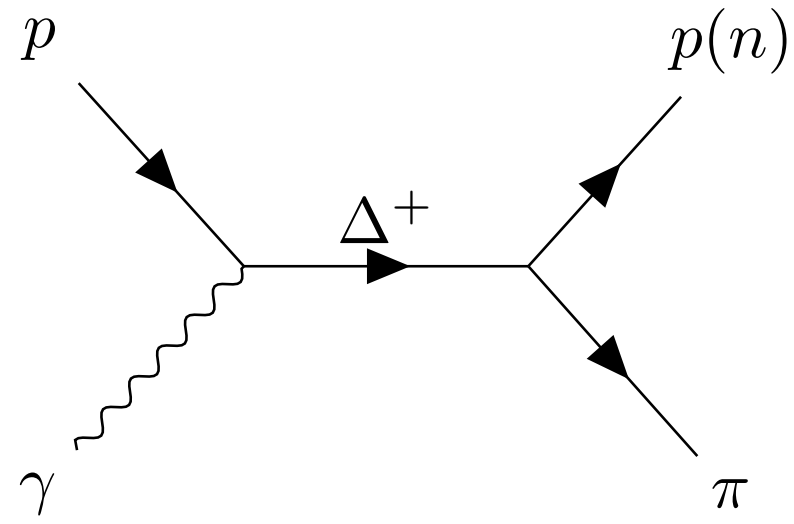

**Figure 3.** Feyman diagram of the most important channel for single pion production.

The main channels for single $\pi^0$ and $\pi^+$ production via the $\Delta^+(1232)$ resonance are:

$$p + \gamma \rightarrow \Delta^+$$

immediately followed by

$$\Delta^+ \rightarrow \begin{cases} n + \pi^+ & b.r. = 1/3 \\ p + \pi^0 & b.r. = 2/3 \end{cases} \tag{7}$$

The cross sections of the $\Delta$ resonance and higher mass $\gamma p$ resonances can be fit with Breit-Wigner type three-parameter distributions of the form

$$\sigma(W) = \frac{W_R^2 \Gamma_R^2 \sigma_{W_R}}{(W^2 - W_R^2)^2 + W_R^2 \Gamma_R^2} \tag{8}$$

where $\sigma_{W_R}$ is the peak in $\sigma(W)$, $W_R$ is the mass of the resonance and $\Gamma_R$ is the width of the resonance.

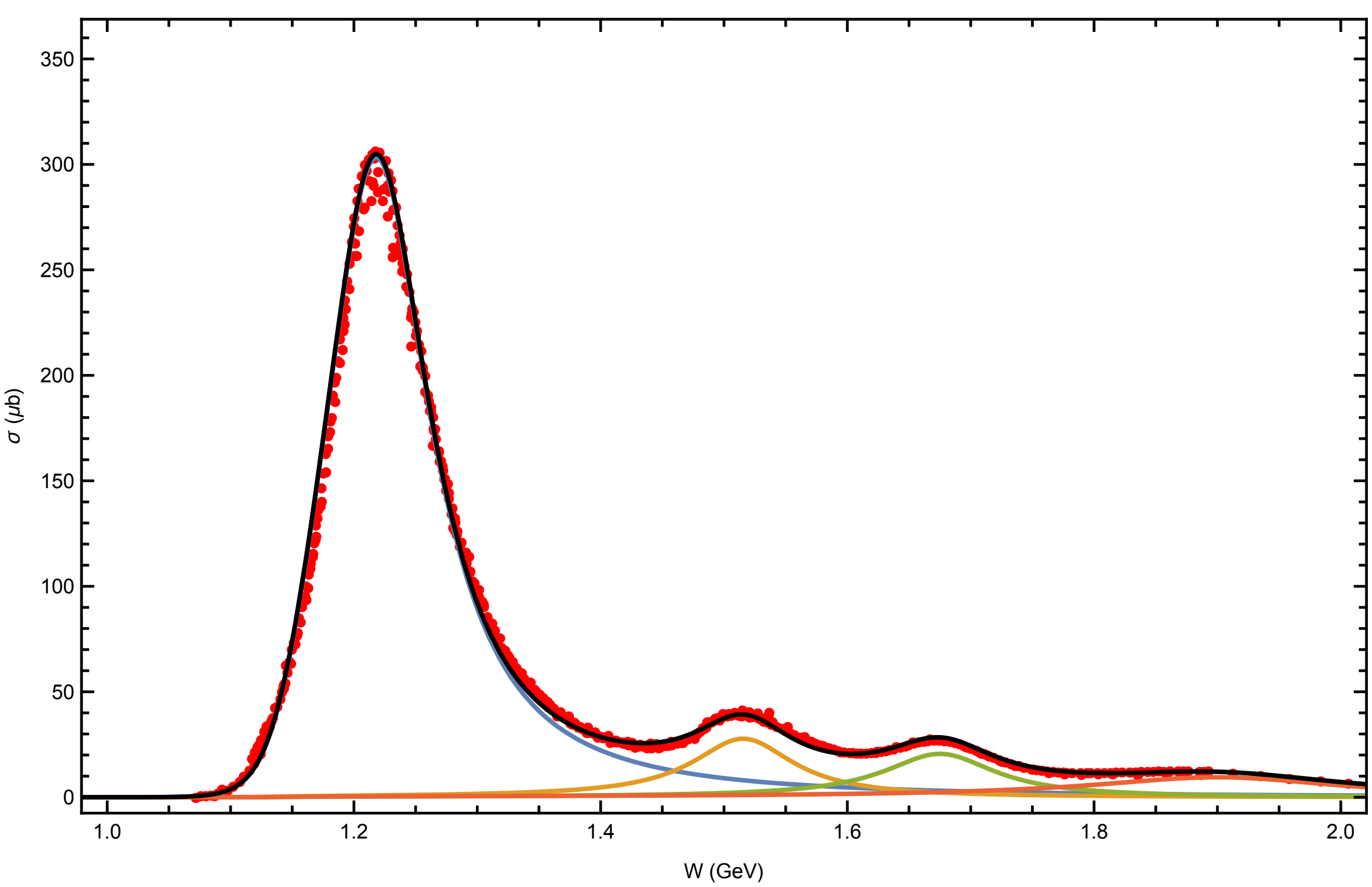

**Figure 4.** A linear plot of the cross sections contributing to single $\pi^0$ production showing the fits to Breit-Wigner type functions and the total cross section compared with the data.

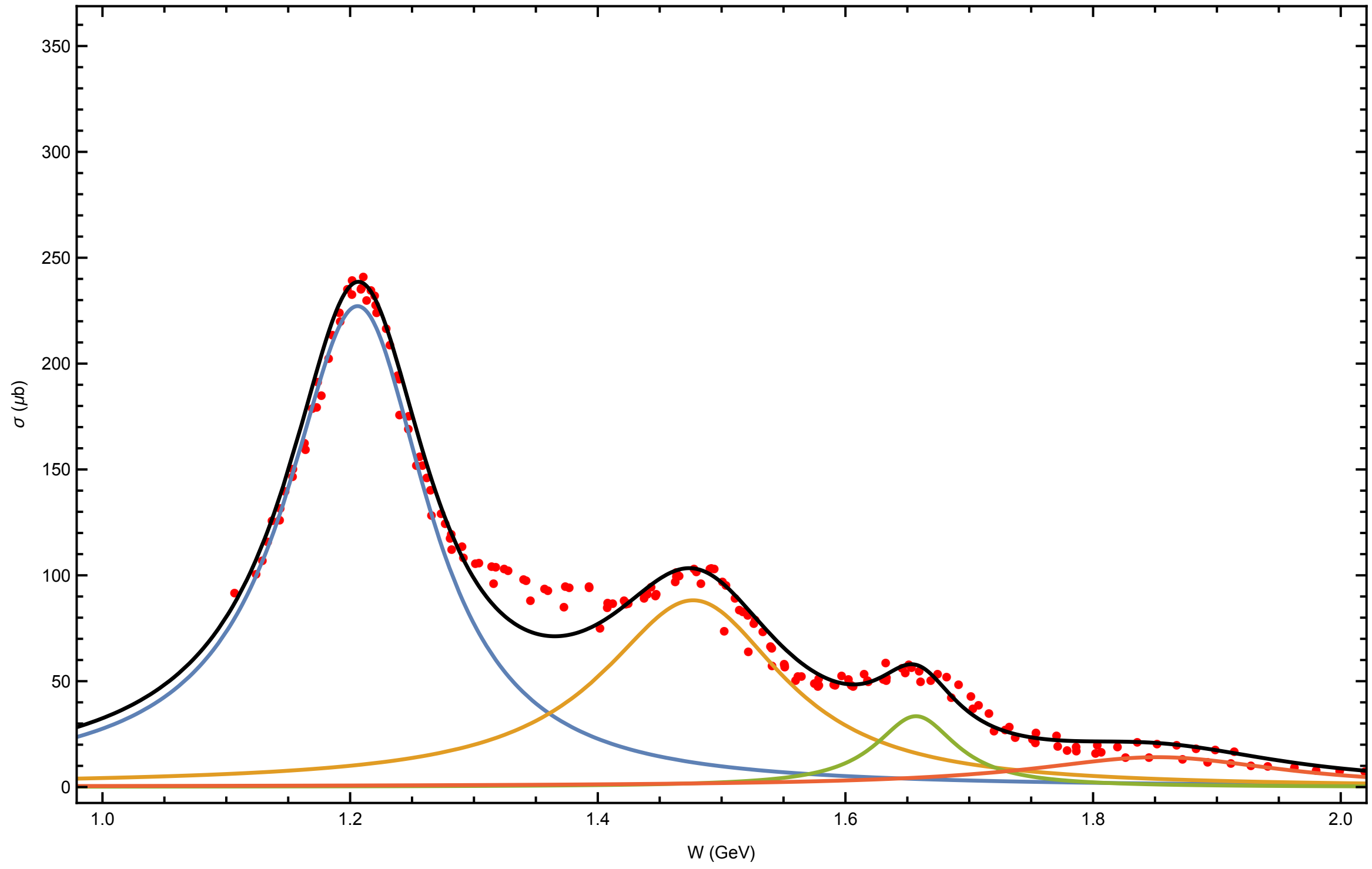

**Figure 5.** A linear plot of the cross sections contributing to single $\pi^+$ production showing the fits to Breit-Wigner type functions and the total cross section compared with the data.

**Table 1.** Breit-Wigner parameter fits for each of the first four resonances for single $\pi^0$ production.

| peak | $W_R(GeV)$ | $\sigma_{W_R}(\mu b)$ | $\Gamma_R(GeV)$ |
|---|---|---|---|
| 1 | 1.217 | 52.9 | 0.110 |
| 2 | 1.516 | 4.35 | 0.100 |
| 3 | 1.675 | 3.55 | 0.110 |
| 4 | 1.900 | 3.70 | 0.250 |

**Table 2.** Breit-Wigner parameter fits for each of the first four resonances for single $\pi^+$ production.

| peak | $W_R(GeV)$ | $\sigma_{W_R}(\mu b)$ | $\Gamma_R(GeV)$ |
|---|---|---|---|
| 1 | 1.206 | 49.7 | 0.140 |
| 2 | 1.477 | 24.8 | 0.180 |
| 3 | 1.655 | 4.80 | 0.090 |
| 4 | 1.852 | 5.00 | 0.250 |

Figure 1 shows the cross section data for $\gamma p$ channels leading to the production of neutral pions and ultimately $\gamma$-rays. The data is given from experiments where high energy $\gamma$-rays interact with protons at rest in the laboratory system.

Figure 2 shows the photoabsorption cross sections for the $\gamma p$ channels leading to charged pion production. The data is given from experiments where high energy $\gamma$-rays interact with protons at rest in the laboratory system. Decay of charged pions then leads to neutrino production.

### 4.2. *Single pion production*

Figure 4 shows the fit obtained by using a combination of several Breit-Wigner type distribution functions with the three parameters shown in eq. (8) to fit the single $\pi^0$ cross section data. Figure 5 shows our fit using three-parameter distribution functions to fit the single $\pi^+$ cross section data.

Note that the data indicate the importance of the $\Delta(1232)$ resonance that appears to contribute to the pion production even at low energies.

Table 1 shows the values obtained for the three-parameter fit of eq.(8), i.e., $W_R$, $\sigma_{W_R}$, and $\Gamma_R$ for the first four resonance peaks obtained by fitting the single $\pi^0$ data shown in Fig. 4.

Table 2 shows the values of $W_R$, $\sigma_{W_R}$, and $\Gamma_R$ for the first four resonance peaks obtained by fitting the single $\pi^0$ data shown in Fig. 5 to eq. (8).

In the data and fits shown in both Fig. 4 and Fig. 5, the first peak (counting from low to high energy) is almost entirely produced by the decay of the $\Delta(1232)$ resonance. This clearly shows the importance of the $\Delta(1232)$ resonance that appears to contribute to single pion production even at low energies. This is particularly important in analyzing models of relativistic protons with a power-law spectrum interacting with photons of the cosmic microwave background and those in the ultraviolet and X-ray range (Greisen 1966; Zatsepin & Kuz'min 1966; Stecker 1968, 1973, 1979).

Both Fig.4 and Fig.5 show at three resonance peaks at higher energy. In Fig. 4 the second peak is primarily a blend of the N(1520) and N(1535) resonance. The third peak is primarily a blend of the N(1650), N(1680) and $\Delta(1700)$ resonances. The fourth peak is primarily a blend of the $\Delta(1920)$ and $\Delta(1950)$ resonances (CB-ELSA Collaboration et al. 2005).

The second, third and fourth peaks are more pronounced in the single $\pi^+$ cross section data than in the single $\pi^0$ cross section data. Also, in particular, the second peak in the single $\pi^+$ production cross section appears at a lower energy ($\sim$1480 MeV) than in the single $\pi^0$ data ($\sim$1530 MeV). The location of this peak could be caused by a blending of the N(1520) and N(1535) resonances with the Roper resonance, N(1440) (Roper 1964). The Roper resonance, which is a minor channel in single $\pi^0$ production, is the only resonance found in the energy range between the $\Delta(1232)$ resonance and the N(1520) resonance.

It should be noted that the positions of the resonance masses can vary from one experiment to another.

The $N(N^*)$ resonances can decay into particles of lighter mass to produce pions. This can include lower mass resonances. For example, we can have three allowed decays of N(1440) Roper resonance:

$$p + \gamma \rightarrow N(1440) \tag{9}$$

immediately followed by

$$N(1440) \rightarrow \begin{cases} N(1440) \rightarrow \Delta + \pi & b.r. \sim 25\% \\ N(1440) \rightarrow N + \pi & b.r. \sim 55\% \\ N(1440) \rightarrow N + \pi + \pi & b.r. \sim 20\% \end{cases} \tag{10}$$

Note that both Fig. 4 and Fig. 5 *automatically take all of the resonances and decays and branching ratios into account*. Furthermore, they break up the cross section data into single $\pi^0$ production and single $\pi^+$ production respectively. The following section treats the double pion decay chains.

### 4.3. *Two pion production*

The second most significant process that leads to the production of charged pions, as illustrated in Figure 2, is the production of $\pi^+\pi^-$. The $\pi^+\pi^-$ production process proceeds through the channels:

$$\gamma + p \rightarrow p + \rho^0 \rightarrow p + \pi^+ + \pi^- \tag{11}$$

$$\gamma + p \rightarrow \Delta^{++} + \pi^- \rightarrow p + \pi^+ + \pi^- \tag{12}$$

$$\gamma + p \rightarrow \Delta^0 + \pi^+ \rightarrow p + \pi^+ + \pi^- \tag{13}$$

At photon energies below 1.0 GeV the cross section for $\gamma + p \rightarrow p + \pi^+ + \pi^-$ is dominated by $\Delta^{++}$ production, viz., $\gamma + p \rightarrow \Delta^{++} + \pi^-$ (Lüke & Söding 1971; Wu et al. 2005). This interaction is followed 100% of the time by the decay $\Delta^{++} \rightarrow p + \pi^+$. At photon energies ranging from 1.0 GeV to 1.4 GeV, a Breit-Wigner distribution provides satisfactory fits to the data, suggesting that these energies are also dominated by the $\Delta^{++}$ resonance. However, it also includes some $\rho$ production above the $\rho$ threshold, followed by the decay $\rho^0 \rightarrow \pi^+ + \pi^-$ with a 100% efficiency (Langgärtner et al. 2001; Hirata et al. 1997). It's worth noting that, unlike the single-pion production process, in the two double-production processes discussed above, the pion energy is approximately evenly distributed between the two pions. The kinematics is therefore greatly simplified as per equation (2).

Figures 1 and 2 both show the total cross section for $\pi^+\pi^0$ photoproduction. The $\pi^+\pi^0$ cross section indicates the presence of the second resonance peak with a chain decay proceeding through a blend of higher $N$ baryon resonances. It also contains a channel for the production $\rho^+$ mesons, $\gamma + p \rightarrow \rho^+ n$ followed by the decay $\rho^+ \rightarrow \pi^+\pi^0$.

### 4.4. *Three pion and gamma-ray line production*

The most important processes for three pion production are the decays

$$\omega \rightarrow \pi^+ + \pi^- + \pi^0 \tag{14}$$

with a branching ratio of 89%,

$$\eta \rightarrow 3\pi^0 \tag{15}$$

and

$$\eta \rightarrow \pi^+ + \pi^- + \pi^0 \tag{16}$$

with a branching ratio of 24%. In these cases we can safely assume that each pion takes away 1/3 of the energy. In addition, the $\eta$ meson can decay into a $\gamma$-ray line

$$\eta \rightarrow \gamma + \gamma \tag{17}$$

with a branching of 28%, with each $\gamma$-ray carrying off 1/2 of the energy of the $\eta$ meson.

## 5. APPLICATION FOR HIGHLY RELATIVISTIC PROTONS IN ASTROPHYSICAL SOURCES

The logarithmic cross section data shown in Figures 1 and 2 show that, for the most part single pion production via resonances dominates, particularly at the lower energies that are most important in considering interactions involving high energy relativistic protons (a.k.a."cosmic rays") that characteristically have a steep ($\propto E^{-n}${where n $\geq$ 2).

For highly relativistic protons ($E \gg m_p, \beta \simeq 1$) interacting with low energy photons, eq.(4) becomes

$$s = m_p + 2\epsilon E_p(1 - cos\theta) \sim 2\epsilon E_p \tag{18}$$

If we set

$$s = M_R^2$$

we can then take as an approximation the spectral relation for the pion flux

$$F(E_\pi) \propto \int F(E_p)\delta(E_p - M_R^2/2\epsilon). \tag{19}$$

where "$\delta$" here designates the well-known $\delta$ function. For a power-law proton spectrum of the form

$$F(E_p) \propto E_p^{-n}$$

eq. (19) gives the result

$$F(E_\pi) \propto M_R^{-2n}. \tag{20}$$

Let us compare the relative fluxes of pions from the $\Delta(1232)$ and the next resonance at N(1500). Denoting the cross sections of these two resonances by $\sigma_1$ and $\sigma_2$ respectively and denoting their masses by $M_1$ and $M_2$ The ratio of of the pion fluxes from their decay that lead to neutrinos and $\gamma$-rays would be given by

$$R \equiv \frac{F_\pi(N(1500))}{F_\pi(\Delta(1232))} = \frac{\sigma_2}{\sigma_1}\left(\frac{M_2}{M1}\right)^{-2n} \tag{21}$$

With $M_2/M_1 \simeq 1.22$ and taking the ratios $\sigma_1$ and $\sigma_2$ for $\pi^0$ and $\pi^+$ from Table 1 and Table 2 respectively, taking $n = 2$, we then find that the ratio of $\gamma$-ray fluxes from the first two resonances is $R_\gamma \sim 0.04$ and the ratio for neutrino fluxes from the first two resonsnces is $R_\nu \sim 0.2$. It follows that for cosmic-ray proton spectra steeper than $E^{-2}$, the $R$ values will be correspondingly smaller.

The contributions of the the other particle species and resonance modes are correspondingly negligible for four reasons: (A) their cross sections are significantly smaller, (B) their production thresholds are higher as discussed previously (See Figure 1 and Figure 2), (C) the steepness of the highly relativistic proton spectra accentuates the threshold effects, and (D) thr effect of multipion decay of the higher mass mesons shown in Figures 1 and 2 splits the energy among the daughter pions produced, as discussed in Sections 4.2 and 4.3 .

## 6. HEAVIER NUCLEI

Photon induced reactions also occur for nucleons of higher mass number. Several experiments have probed the nuclear response to photons with total photon absorption measurements having been made from nuclei with mass numbers ranging through to $^{238}$U. In general, nuclear cross sections are very similar to those resulting from $\gamma p$ interactions, when scaled by the atomic mass number to some power $\leq 1$, with the exception that the cross section data for nuclei heavier than hydrogen show no resonance structure in the resonance regions higher than the $\Delta(1232)$ resonance. The shape of the $\Delta(1232)$ resonance is also broadened. These differences from $\gamma p$ cross section measurements can be explained by taking into account the cooperative effect of the collision broadening and the interference in the two-pion photoproduction process inside the nuclei (Hirata et al. 2002).

It follows from equation (1) that the threshold energy for photopion production off helium nuclei is $\sim 0.137$ GeV for $\pi^0$ and $\sim 0.143$ GeV for $\pi^+$. The cross section for photons of energies $200 - 800$ MeV interacting with He nuclei at rest for both $\pi^0$ and $\pi^\pm$ production have been taken from the measurements of MacCormick et al. (1997). In $\gamma He$ interactions low laboratory photon energies single pion production from the $\Delta(1232)$ resonance decay channel is again dominant.

Given the astrophysical abundances of the elements, only helium makes any significant contribution to photomeson production in astrophysical sources and only in the region dominated by the $\Delta(1232)$ resonance. By incorporating an additional cosmological helium component that constitutes 25% by mass (Kolb & Turner 2019), and by using the cross section data for helium (MacCormick et al. 1997), one can estimate the magnitudes of $\gamma$-ray and neutrino fluxes from photoproduction off He as compared with those from $\gamma p$ interactions. Using the proton and He cross sections at the $\Delta(1232)$ resonance energy to give an estimate of the He contribution, one finds that the flux from $\gamma$He interactions is $\sim$10% of that from $\gamma p$ interactions.

## 7. DISCUSSION OF RESULTS

In this paper, our aim has been to reconsider the fundamental physics of the photoproduction processes that are of astrophysical significance. We have shown how and why there is a fundamental difference between the physics of $pp$ interactions and that of $\gamma p$ interactions. In high energy $pp$ interactions the multipion production (pionization) that results from QCD physics plays a significant role in $\gamma$-ray production (Stecker 1970, 1971). However, in $\gamma p$ interactions it is the production of single particles and resonances that plays the major role. This is evidenced by the library of cross section data compiled here. We delineate a list of our new conclusions as follows:

1. By use of this collection of empirical cross section data one can determine the relative role of particle production channels in $\gamma p$ interactions leading to neutrino and $\gamma$-ray production in various astrophysical contexts. By compiling the cross sections for all of the channels that result in the production of $\gamma$-rays and neutrinos as a result of $\gamma p$ interactions, we have determined that only the single pion resonance channels are significant for astrophysical considerations. The cross section data compiled here show that, contrary to the assertion of the SOPHIA collaboration (Mücke et al. 2000), multiparticle production channels are not significant in producing $\gamma$-rays in $\gamma p$ interactions.

2. We further find that two and three pion production is actually only the result of decay chains following from single particle photoproduction. There is no monotonic increase in multiplicity with energy as in the case of $pp$ interctions. This is especially important when one takes account of the steep proton spectra, e.g., $E^{-2.5}$ or steeper, expected for astrophysical sources of high energy protons.

3. As shown in our Figures 4 and 5, there are significant differences between the production of charged and neutral pions (see point 5). This leads to significant differences between $\gamma$-ray and neutrino production that are not accounted for in SOPHIA-based models.

4. As opposed to the case of pp interactions, and some speculation about $\gamma - p$ interactions, owing to the small production cross sections and high threshold energy for kaons (see Section 3.) $K$ production is negligible compared with pion production. This is abundantly clear from the data shown in Figs 1 and 2.

5. As expected, among the single pion production channels, the $\Delta(1232)$ channel dominates, even among the higher resonance channels. However, in the case of charged pion decay we find that the second resonance peak, made up of a blend on $N^*$ resonances, can give an additional contribution of as much as 20% to the resulting production of neutrinos. On the contrary, we find that the second resonance makes a neglible contribution to neutral pion production and subsequently to$\gamma$-ray production.

6. For the first time, we have considered photoproduction of particles off heavier nuclei. As expected, only helium is important for astrophysical sources, given the relative cosmological abundances of the heavier elements. For helium, as for hydrogen, photoproduction is important only the region dominated by the Delta(1232) resonance. We find that the $\gamma$-ray and neutrino fluxes resulting from $\gamma He$ interactions is $\sim$10% of that from $\gamma p$ interactions.

7. The use of a delta function approximation to simulate the $\Delta(1232)$ production of $\gamma$-rays and neutrinos Stecker (1973, 1979); Halzen & Kheirandish (2023) is quite adequate for most calculations, given the present state of $\gamma$-ray and neutrino astronomy. However, a detailed knowledge the physics of photopion production may be more importance in the future as given advances in observational instrumentation and source modeling.

# 8. APPENDIX:PARTICLE SPECTRA

## 8.1. *Gamma-ray spectra*

Given a production function of neutral pions, $f(E_\pi)$, the $\gamma$-ray source spectrum that results is given by

$$q(E_\gamma) = 2\int_{\xi}^{\infty} dE_\pi f(E_\pi)(E_\pi^2 - m_\pi^2)^{-1/2} \tag{22}$$

where $\xi = E_\gamma + m_\pi^2/4E_\gamma$ (Stecker 1971).

## 8.2. *Neutrino spectra*

Neutrinos are produced in two stages of pion decay. The first stage of the decay

$$\pi \rightarrow \mu + \nu \tag{23}$$

is a two particle decay. So as in eq. (2) we find that the neutrino energy in the rest system of the pion is

$$E_\nu^* = E_\pi^* \left(\frac{m_\pi^2 - m_\mu^2}{2m_\pi}\right). \tag{24}$$

(see kinematics in Stecker 1971, chap. 1). It follows from eq. (24) that the maximum energy of the neutrino is then given by

$$E_\nu^{max} \;=\; \left(\frac{m_\pi^2 - m_\mu^2}{m_\pi^2}\right) E_\pi \;=\; 0.427 E_\pi \;\equiv\; \eta E_\pi \tag{25}$$

and the minimum energy is approximately zero. The average energy for isotropic decay is then $< E_\nu \simeq 0.21 E_\pi$. This is close to the 25% expected if the pion decays into four very light fermions. Then, for isotropic decay of a relativistic pion, the source function for neutrinos from the first stage of the decay only is

$$q(E_\nu) \simeq \int_{\eta}^{\infty} \frac{dE_\pi f(E_\pi)}{\eta E_\pi}. \tag{26}$$

The second stage of the process is the decay of the $\mu$ ,

$$\mu^+ \rightarrow \mu_\nu^+ + \bar{\nu}_e + e^+ \tag{27}$$

and the equivalent process for $\mu^-$ decay. In these decays, since the three leptons produced are much lighter than the muon, each lepton will get an average of a third of the muon energy or about 26% of the pion energy. In fact, one can assume that all of the neutrinos each get about a quarter of the pion energy.